%% file: main.tex
\documentclass[sigconf]{acmart}

\makeatletter
\def\@ACM@checkaffil{
    \if@ACM@instpresent\else
    \ClassWarningNoLine{\@classname}{No institution present for an affiliation}%
    \fi
    \if@ACM@citypresent\else
    \ClassWarningNoLine{\@classname}{No city present for an affiliation}%
    \fi
    \if@ACM@countrypresent\else
        \ClassWarningNoLine{\@classname}{No country present for an affiliation}%
    \fi
}
\makeatother

\AtBeginDocument{%
  \providecommand\BibTeX{{%
    \normalfont B\kern-0.5em{\scshape i\kern-0.25em b}\kern-0.8em\TeX}}}

\acmConference[ICSE 2024]{46th International Conference on Software Engineering}{April 2024}{Lisbon, Portugal}

\input{macro}

\begin{document}


\title{Testing the Limits: Unusual Text Inputs Generation for Mobile App Crash Detection with Large Language Model}

\author{Zhe Liu$^{1}$,Chunyang Chen$^2$, Junjie Wang$^{1,*}$, Mengzhuo Chen$^{1}$, Boyu Wu$^{1}$, Zhilin Tian$^{1}$, \\ Yuekai Huang$^{1}$, Jun Hu$^{1}$, Qing Wang$^{1,*}$}
\affiliation{
  \position{$^1$State Key Laboratory of Intelligent Game, Beijing, China}
  \department{Institute of Software Chinese Academy of Sciences, Beijing, China; \\
  University of Chinese Academy of Sciences, Beijing, China; $^*$Corresponding author\\
  $^2$Monash University, Melbourne, Australia;
  }
}
\email{liuzhe181@mails.ucas.ac.cn, Chunyang.chen@monash.edu, junjie@iscas.ac.cn, wq@iscas.ac.cn}

\begin{abstract}
Mobile applications have become a ubiquitous part of our daily life, providing users with access to various services and utilities. 
Text input, as an important interaction channel between users and applications, plays an important role in core functionality such as search queries, authentication, messaging, etc.
However, certain special text (e.g., -18 for Font Size) can cause the app to crash, and generating diversified unusual inputs for fully testing the app is highly demanded.
Nevertheless, this is also challenging due to the combination of explosion dilemma, high context sensitivity, and complex constraint relations. 
This paper proposes {\tool} which leverages the LLM to automatically generate unusual text inputs for mobile app crash detection. 
It formulates the unusual inputs generation problem as a task of producing a set of test generators, each of which can yield a batch of unusual text inputs under the same mutation rule. 
In detail, {\tool} leverages LLM to produce the test generators together with the mutation rules serving as the reasoning chain, and utilizes the in-context learning schema to demonstrate the LLM with examples for boosting the performance. 
{\tool} is evaluated on 36 text input widgets with cash bugs involving 31 popular Android apps, and results show that it achieves 78\% bug detection rate, with 136\% higher than the best baseline. 
Besides, we integrate it with the automated GUI testing tool and detect 37 unseen crashes in real-world apps from Google Play.

\end{abstract}


\maketitle



\input{sec/introduction}

\input{sec/motivation}

\input{sec/approach}

\input{sec/experiment}

\input{sec/result}

\input{sec/discussion}

\input{sec/related}
\input{sec/conclusion}




\bibliographystyle{ACM-Reference-Format}

\bibliography{reference}

\end{document}

%% file: macro.tex
\usepackage{soul}
\usepackage{float}
\usepackage{threeparttable}
\usepackage{booktabs}
\usepackage{multirow}
\usepackage{epstopdf}
\usepackage{hyperref}
\usepackage{listings}
\usepackage{graphicx}
\usepackage{subcaption}
\usepackage{paralist}
\usepackage{ragged2e}
\usepackage{enumitem}

\usepackage[normalem]{ulem} 
\newcommand\hlg{\bgroup\markoverwith
  {\textcolor{green!10}{\rule[-.5ex]{2pt}{2.5ex}}}\ULon}
\newcommand\hlb{\bgroup\markoverwith
  {\textcolor{blue!10}{\rule[-.5ex]{2pt}{2.5ex}}}\ULon}
\newcommand\hlr{\bgroup\markoverwith
  {\textcolor{red!10}{\rule[-.5ex]{2pt}{2.5ex}}}\ULon}

\usepackage[framemethod=TikZ]{mdframed}
\usepackage{tikz}
\usepackage{pgfplots}
\usetikzlibrary{pgfplots.statistics,calc}
\usepackage{color}
\usepackage{xcolor}
\usepackage{array}
\usepackage{amsmath}
\usepackage{centernot}
\usepackage{url}
\usepackage{verbatim}
\usepackage{wrapfig}
\usepackage{tabularx}
\usepackage{bbding}
\usepackage{pifont}
\usepackage{wasysym}
\usepackage{amssymb}

\newcommand{\tool}{InputBlaster}

\makeatletter  
\newif\if@restonecol  
\makeatother  
\usepackage[linesnumbered,ruled,vlined]{algorithm2e}
\usepackage{algpseudocode}  
\usepackage{amsmath}  

\usepackage[english]{babel}

%% file: sec/introduction.tex
\section{Introduction}
\label{sec_introduction}

Mobile applications (apps) have become an indispensable component of our daily lives, enabling instant access to a myriad of services, information, and communication platforms. The increasing reliance on these applications necessitates a high standard of quality and performance to ensure user satisfaction and maintain a competitive edge in the fast-paced digital landscape. The ubiquity of mobile applications has led to a constant need for rigorous testing and validation to ensure their reliability and resilience against unexpected user inputs. 

Text input plays a crucial role in the usability and functionality of mobile applications, serving as a primary means for users to interact with and navigate these digital environments~\cite{liu2022fill,liu2017automatic}. From search queries and form submissions to instant messaging and content creation, text input is integral to the core functionality of numerous mobile applications across various domains. The seamless handling of text input is essential for delivering a positive user experience, as it directly impacts the ease of use, efficiency, and overall satisfaction of the users.

Given the unexpected input, the program might suffer from memory leakage, data corruption, falling into the dead loop, resulting in the application stuck, crash, or other serious issues~\cite{trinh2014s3,he2020textexerciser,chen2022solving,holik2017string}.
Even worse, these buggy texts can only demonstrate a tiny difference from the normal text, or they themselves are normal text in other contexts, which makes the issue easily occur and difficult to spot. 
There has been a fair amount in the news about the crash 
of iOS and Android systems caused by a special text input~\cite{inputissue}, which has greatly affected people's daily lives. 
For example, in July 2020, a specific character of the Indian language caused iOS devices constantly crash.
It has affected a wide range of iOS applications, including iMessage, WhatsApp, and Facebook Messenger~\cite{inputissue-2}, and as long as certain text inputs contain the character, these apps would crash. 


Taken in this sense, automatically generating unusual inputs for fully testing the input widgets and uncovering bugs is highly demanded.
Existing automated GUI testing techniques focus on generating the valid text input for passing the GUI page and conducting the follow-up page exploration \cite{liu2022fill,liu2017automatic,trinh2014s3,he2020textexerciser,anand2012automated,arnatovich2018mobolic,sunman2022automated}, e.g., QTypist~\cite{liu2022fill} used GPT-3 to generate semantic input text to improve the coverage of the test.
They could not be easily adapted to this task, since the unusual inputs can be more diversified and follow different rationales from the valid inputs. 
There are also studies targeting at generating strings that violate the constraints (e.g., string length) with heuristic analysis or finite state automaton techniques~\cite{trinh2017model,liang2014dpll,li2013pass}.
Yet they are designed for specific string functions like concatenation and replacement, and could not be generalized in this task.

Nevertheless, it is very challenging for the automatic generation of diversified unusual inputs.
The first challenge is the combination explosion. There can be numerous input formats including text, number, date, time, currency, and innumerable settings, e.g., different character sets, languages and text lengths, which makes it quite difficult if not impossible to enumerate all these variants. 
The second challenge is context sensitivity. The unusual inputs should closely relate to the context of the input widgets to effectively trigger the bug, e.g., a negative value for font size (as shown in Figure \ref{fig:Example}), an extremely large number to potentially violate the widget for people's height. 
The third challenge is the constraint relation within and among the input widgets. 
The constraints can be that a widget only accepts pure numbers (without characters), or the sum of item values smaller/bigger than the total (as shown in Figure \ref{fig:Example}), which requires an exact understanding of the related widgets and these constraints so as to generate targeted variation.
What's more difficult is that certain constraints only appear when interacting with the apps (i.e., dynamic hints in terms of the incorrect texts), and static analysis cannot capture these circumstances.

\input{figure/Example}



Large Language Models (LLMs)~\cite{attention,brown2020GPT3,chowdhery2022palm,zhang2022opt,schulman2022chatgpt}
trained on ultra-large-scale corpus have exhibited promising performance in a wide range of tasks. ChatGPT\cite{schulman2022chatgpt}, developed by OpenAI, is one such LLM with an impressive 175 billion parameters, trained on a vast dataset. Its ability to comprehend and generate text across various domains is a testament to the potential of LLMs in interacting with humans as knowledgeable experts. The success of ChatGPT is a clear indication that LLMs can understand human knowledge and can do well in providing answers to various questions.



Inspired by the fact that the LLM has made outstanding progress in email reply, abstract extraction, etc.~\cite{brown2020GPT3,laskin2020reinforcement,yang2022empirical,chen2020big}, 
we propose an approach, {\tool}\footnote{Our approach is named as {\tool} considering it likes a blaster which ignites the following production of the unusual inputs.}, to automatically generate the unusual text inputs with LLM which uncover the bugs\footnote{Note that, like existing studies \cite{DroidBot,li2019humanoid,pan2020reinforcement}, this paper focuses on the crash bug, which usually causes more serious effects and can be automatically observed, and we interchangeably use the term bug and crash.} related to the text input widgets.
Instead of directly generating the unusual inputs by LLM which is of low efficiency, we formulate the unusual inputs generation problem as a task of producing a set of test generators (a code snippet), each of which can yield a batch of unusual text inputs under the same mutation rule (i.e., insert special characters into a string), as demonstrated in Figure \ref{fig:in-context-learning} \ding{176}.

To achieve this, {\tool} leverages LLM to produce the test generators together with the mutation rules which serve as the reasoning chains for boosting the performance. 
In detail, {\tool} first leverages LLM to generate the valid input which can pass the GUI page and serves as the target for the follow-up mutation (Module 1).
Based on it, it then leverages LLM to produce mutation rules, and asks the LLM to follow those mutation rules and produce the test generator, each of which can yield a batch of unusual text inputs (Module 2).
To further boost the performance, we utilize the in-context learning schema to demonstrate the LLM with useful examples from online issue reports and historical running records (Module 3).  


To evaluate the effectiveness of {\tool}, we carry out experiments on 36 text input widgets with cash bugs involving 31 popular Android apps in Google Play.
Compared with 18 common-used and state-of-the-art baselines, {\tool} can achieve more than 136\% boost in bug detection rate compared with the best baseline, resulting in 78\% bugs being detected. 
In order to further understand the role of each module and sub-module of the approach, we conduct ablation experiments to further demonstrate its effectiveness.
We also evaluate the usefulness of {\tool} by integrating it with the automated GUI testing tool and detecting unseen crash bugs in real-world apps from Google Play. 
Among 131 apps, {\tool} detects 37 new crash bugs with 28 of them being confirmed and fixed by developers, while the remaining are still pending. 

The contributions of this paper are as follows:
\begin{itemize}



\item We are the first to propose a novel LLM-based approach {\tool} for the automatic generation of unusual text inputs for mobile app testing.



\item We conduct the first empirical categorization of the constraint relationships within and among text input widgets, which provides clues for the LLM in effective mutation, and facilitates the follow-up studies on this task.  

\item We carry out the effectiveness and usefulness evaluation of {\tool}, with a promising performance largely outperforming baselines and 37 new detected bugs. 
 
\end{itemize}

%% file: figure/Example.tex
\begin{figure}[htb]
\centering
\includegraphics[width=8.6cm]{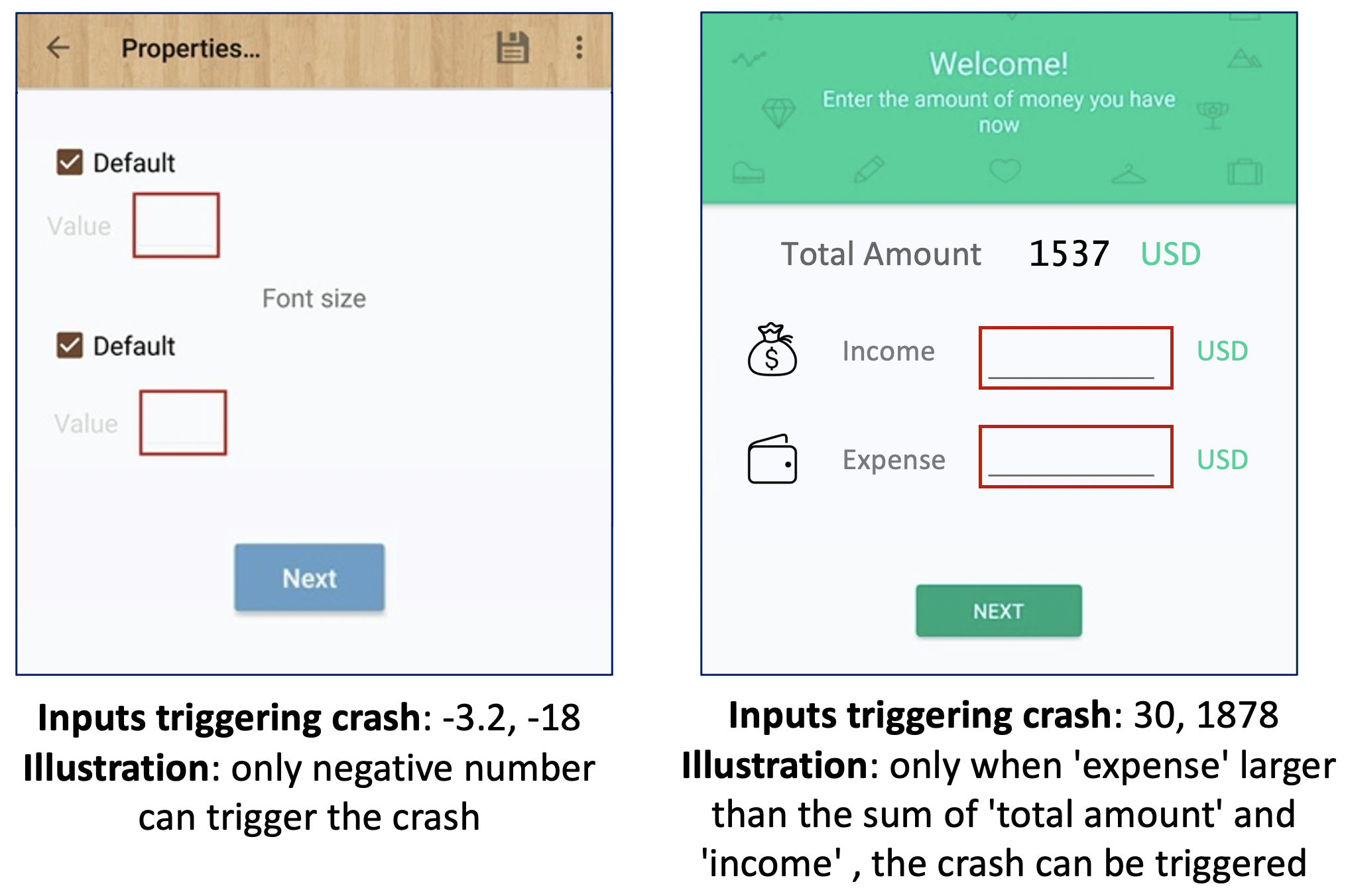}
\caption{Example bugs triggered by unusual inputs.}
\label{fig:Example}
\vspace{-0.1in}
\end{figure}

%% file: sec/motivation.tex
\section{motivational study and background}
\label{sec_motivation}
To better understand the constraints of text inputs in real-world mobile apps, we carry out a pilot study to examine their prevalence. 
We also categorize the constraints, to facilitate understanding and the design of our approach for generating unusual inputs violating the constraints. 

\subsection{Motivational Study}
\label{sec_motivation_data_collection}
\subsubsection{\textbf{Data Collection}}
The dataset is collected from one of the largest Android GUI datasets Rico~\cite{Rico}, which has a great number of Android GUI screenshots and their corresponding view hierarchy files~\cite{liu2020owl,liu2022Nighthawk}.
These apps belong to diversified categories such as news, entertainment, medical, etc. 
We analyze the view hierarchy file according to the package name and extract the GUI page belonging to the same app. 
A total of 7,136 apps with each having more than 3 GUI  pages are extracted.
For these apps, we first randomly select 136 apps with 506 GUI pages and check their text inputs through view hierarchy files. 
We summarize a set of keywords that indicate the apps have text inputs widgets~\cite{Textinput}, 
e.g., \textit{EditText, hint-text, AutoCompleteTextView, etc}.
We then use these keywords to automatically filter the view hierarchy files from the remaining 7,000 apps, and obtain 5,761 candidate apps with at least one potential text input widget.
Four authors then manually check them to ensure that they have text inputs until a consensus is reached.
In this way, we finally obtain 5,013 (70.2\%) apps with at least one text input widget, and there are 3,723 (52.2\%) apps having two or more text input widgets. Please note that there is no overlap with the evaluation dataset.




\subsubsection{\textbf{The Constraint Categories of Text Inputs}}
\label{sec_motivation_types_UIs}
We randomly select 2000 apps with text inputs and conduct manual categorization to derive the constraint types of input widgets. 
Following the open coding protocol~\cite{seaman1999qualitative}, two authors 
individually examine the content of the text input, including the app name, activity name, input type and input content.
Then each annotator iteratively merges similar codes, and any disagreement of the categorization will be handed over to the third experienced researcher for double checking. 
Finally, we come out with a categorization of the constraints within (intra-widget) and among the widgets (inter-widget), with details summarized in Figure \ref{fig:motivation}.

\textbf{Intra-widget constraint.}  Intra-widget constraints depict the requirements of a single text input, e.g., a widget for a human's height can only input the non-negative number.
There are explicit and implicit sub-types. 
The former accounts for 63\%, which manifests as the requirement to display input directly on the GUI page. 
And the latter account for 37\%, mainly manifested as the feedback when incorrect text input is received, e.g., after inputting a simple password, the app would remind the user ``at least one upper case character (A-Z) is required'' as demonstrated in Figure \ref{fig:motivation}.

\textbf{Inter-widget constraint.} 
Inter-widget constraints depict the requirements among multiple text input widgets on a GUI page, for example, the diastolic pressure should be less than systolic pressure as shown in Figure \ref{fig:motivation}.


\textbf{Summary.} As demonstrated above, the text input widgets are quite common in mobile apps, e.g., 70.2\% apps with at least one such widget. 
Furthermore, considering the diversity of inputs and contexts, it would require significant efforts to manually build a complete set of mutation rules to fully test an input widget, and the automated technique is highly demanded. 
This confirms the popularity of text inputs in mobile apps and the complexity of it for full testing, which motivates us to
automatically generate a batch of unusual text inputs for effective testing and bug detection.

\subsection{Background of LLM and In-context Learning}
\label{sec_motivation_Large Language Model}
The target of this work is to generate the input text, and the Large Language Model (LLM) trained on ultra-large-scale corpus can understand the input prompts (sentences with prepending instructions or a few examples) and generate reasonable text.
When pre-trained on billions of samples from the Internet, recent LLMs (like ChatGPT~\cite{schulman2022chatgpt}, GPT-3~\cite{brown2020GPT3} and T5~\cite{DBLP:journals/jmlr/RaffelSRLNMZLL20}) encode enough information to support many natural language processing tasks~\cite{yang2022empirical,lucy2021gender,sharples2022automated}. 

Tuning a large pre-trained model can be expensive and impractical for researchers, especially when limited fine-tuned data is available for certain tasks. In-context Learning (ICL) \cite{brown2020language,min2022rethinking,garg2022can} offers a new alternative that uses Large Language Models to perform downstream tasks without requiring parameter updates. It leverages input-output demonstration in the prompt to help the model learn the semantics of the task. This new paradigm has achieved impressive results in various tasks, including code generation and assertion generation.
\input{figure/motivation}

\input{figure/overview}

%% file: figure/motivation.tex
\begin{figure}[htb]
\centering
\includegraphics[width=8.3cm]{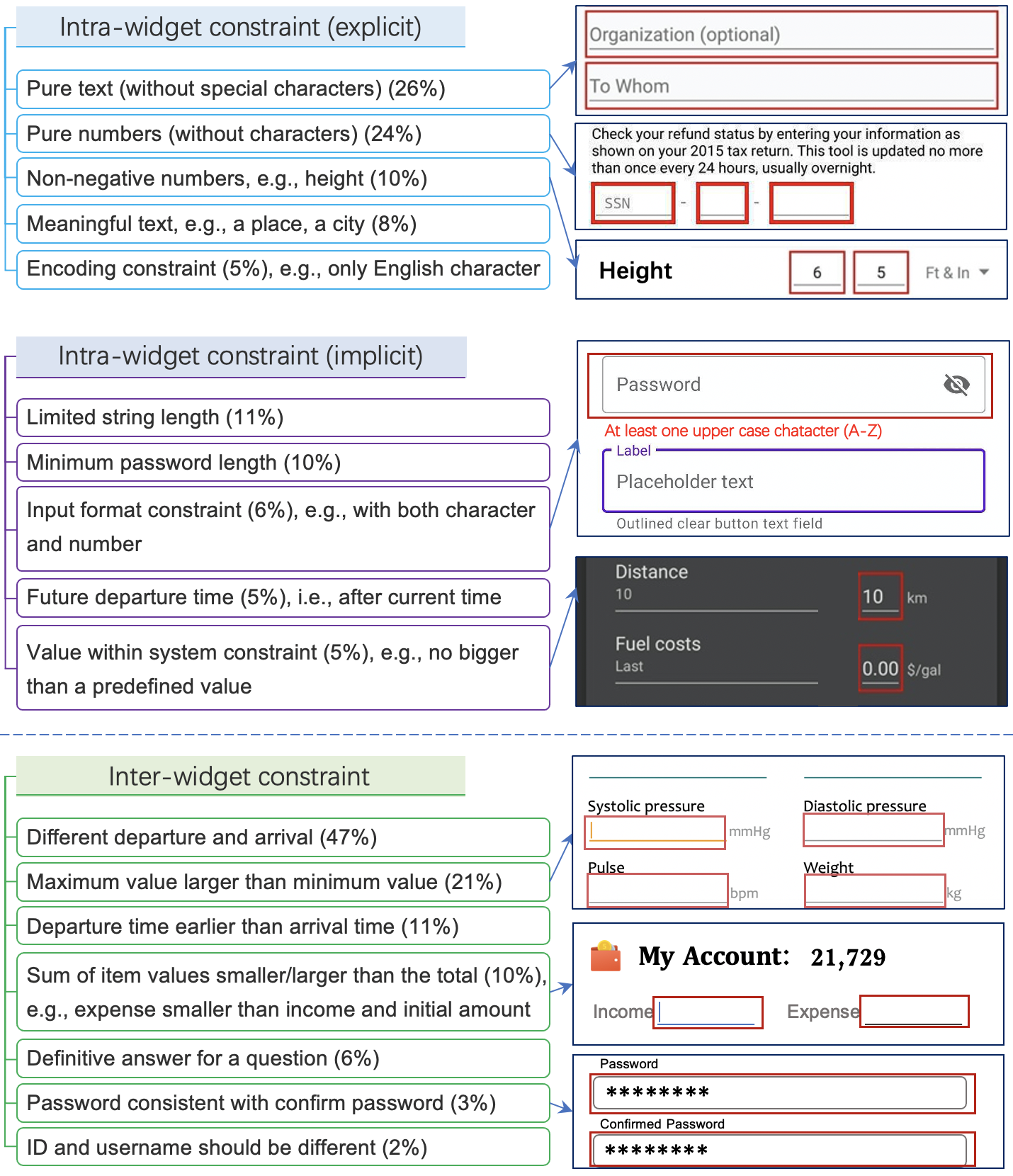}
\vspace{-0.1in}
\caption{The category of constraints.}
\label{fig:motivation}
\vspace{-0.15in}
\end{figure}

%% file: figure/overview.tex
\begin{figure*}[t]
\centering
\includegraphics[width=17.3cm]{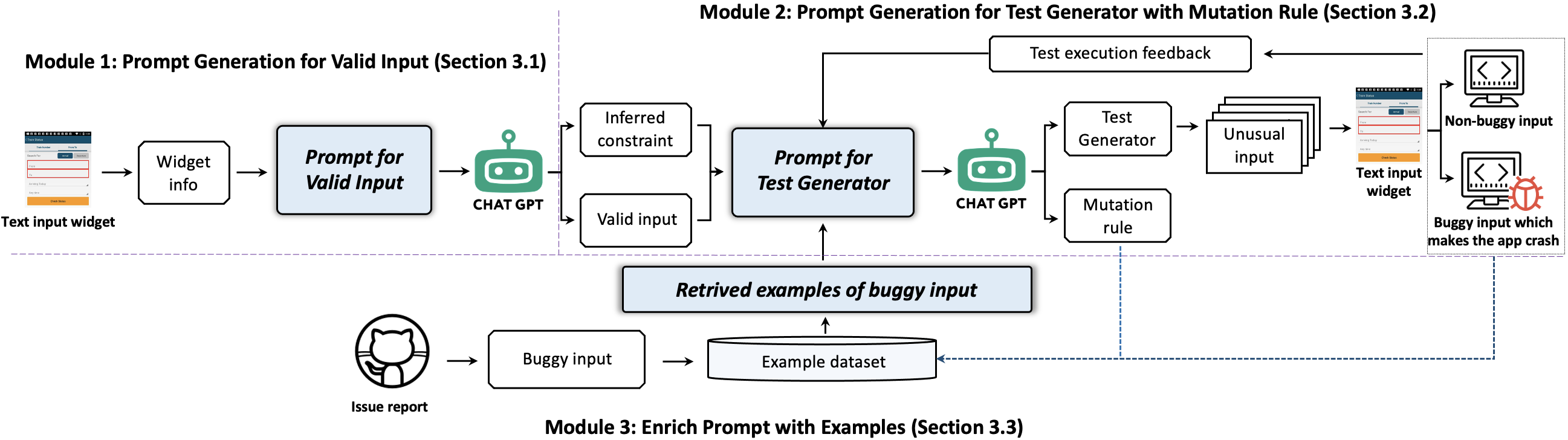}
\vspace{-0.1in}
\caption{Overview of {\tool}.}
\label{fig:overview}
\vspace{-0.15in}
\end{figure*}

%% file: sec/approach.tex
\section{Approach}
\label{sec_approach}

This paper aims at automatically generating a batch of unusual text inputs which can possibly make the mobile apps crash. 
The common practice might directly produce the target inputs with LLM as existing studies in valid input generation~\cite{liu2022fill} and fuzzing deep learning libraries~\cite{deng2022fuzzing,deng2023large}. 
Yet, this would be quite inefficient for our task, because each interaction with the LLM requires a few seconds waiting for the response and consumes lots of energy. 
Instead, this paper proposes to produce the test generators (a code snippet) with LLM, each of which can generate a batch of unusual text inputs under the same mutation rule (e.g., insert special characters into a string), as demonstrated in Figure \ref{fig:in-context-learning} \ding{176}.

To achieve this, we propose {\tool} which leverages LLM to produce the test generators together with the mutation rules which serve as the reasoning chains for boosting the performance, and each test generator then automatically generates a batch of unusual text inputs, as shown in Figure \ref{fig:overview}. 
In detail, given a GUI page with text input widgets and its corresponding view hierarchy file, we first leverage LLM to generate the valid text input which can pass the GUI page (Sec \ref{subsec_approach_Prompt_Generation}).
We then leverage LLM to produce the test generator which can generate a batch of unusual text inputs, and simultaneously we also ask the LLM to output the mutation rule which serves as the reasoning chain for guiding the LLM in making the effective mutations from valid inputs (Sec \ref{subsec_approach_Test_Prompt_Generation}).   
To further boost the performance, we utilize the in-context learning schema to provide useful examples when querying the LLM, from online issue reports and historical running records (Sec \ref{subsec_approach_Prompt-tuning}). 

\input{figure/in-context-learning}



\subsection{Prompt Generation for Valid Input}
\label{subsec_approach_Prompt_Generation}

{\tool} first leverages LLM to generate the valid input which will serve as the target towards which the following mutation can be conducted. 
The context information relates to the input widgets and its belonged GUI page can provide important clues about what the valid input should be, therefore we input this information into LLM (in Section \ref{subsubsec_approach_Context_Extraction}).
In addition, we also include the dynamic feedback information when interacting with the input widgets (in Section \ref{subsubsec_approach_dynamic_hint_Extraction}), and the constraint categories we summarized in the previous section (in Section \ref{subsubsec_approach_candidate_constraint}) to improve the performance. 
Furthermore, besides the valid text input, we also ask LLM to output its inferred constraints for generating the valid input which will facilitate the approach to generating the mutation rules in the next section. 
We summarize all the extracted information with examples in Table \ref{tab:approach-rule}.


\subsubsection{\textbf{Context Extraction}}
\label{subsubsec_approach_Context_Extraction}

The context information is extracted from the view hierarchy file, which is easily obtained by automated GUI testing tools~\cite{machiry2013dynodroid,mao2016sapienz,gu2019practical,su2017guided}. 
As shown in Table \ref{tab:approach-rule}, we extract the text-related field of the input widget which indicates how the valid input should be. 
In detail, we extract the ``hint text'', ``resource id'', and `text' fields of the input widget, and utilize the first non-empty one among the above three fields.

We also extract the activity name of the GUI page and the mobile app name, and this global context further helps refine the understanding of the input widget.
In addition, we extract the local context of the input widget (i.e., from nearby widgets) to provide thorough viewpoints and help clarify the meaning of the widget. 
The candidate information source includes the parent node widgets, the leaf node widget, widgets in the same horizontal axis, and fragment of the current GUI page.
For each information source, we extract the ``text'' field (if it is empty, use the ``resource-id'' field), and concatenate them into the natural-language description with the separator (`;').

\subsubsection{\textbf{Dynamic Hint Extraction}}
\label{subsubsec_approach_dynamic_hint_Extraction}

When one inputs an incorrect text into the app, there are some feedbacks (i.e., dynamic hints) related to the inputs, e.g., the app may alter the users that the password should contain letters and digits. 
The dynamic hint can further help LLM understand what the valid input should look like. 


We extract the dynamic hints via differential analysis which compares the differences of the GUI page before and after inputting the text, and extracts the text field of the newly emerged widgets (e.g., a popup window) in the later GUI page, with examples shown in Figure \ref{fig:motivation}.
We also record the text input which makes the dynamic hint happens, which can help the LLM to understand the reason behind it.   

\subsubsection{\textbf{Candidate Constraints Preparation}}
\label{subsubsec_approach_candidate_constraint}

Our pilot study in Section \ref{sec_motivation_types_UIs} summarizes the categories of constraints within and among the widgets. 
The information can provide direct guidance for the LLM in generating the valid inputs, for example, the constraint explicitly requires the input should be pure text (without special characters). 
We provide this list of all candidate constraints described in natural language as in Section \ref{sec_motivation_types_UIs} to the LLM.


\input{tab/approach-rule}

\subsubsection{\textbf{Prompt Generation}}
\label{subsubsec_approach_Prompt_Generation}

With the extracted information, we use three kinds of information to generate prompts for inputting into the LLM, as shown in Table \ref{tab:approach-rule}.
Generally speaking, it first provides the context information and the dynamic hints (if any) of the input widgets, followed by the candidate constraints, and then queries the LLM for the valid input. 
Due to the robustness of LLM, the generated prompt sentence does not need to fully follow the grammar. 

After inputting the prompt, the LLM will return its recommended valid text input and its inferred constraints, as demonstrated in Figure \ref{fig:in-context-learning} \ding{173}.
We then input it into the widget, and check whether it can make the app transfer to the new GUI page (i.e., valid input). 
If the app fails to transfer, we iterate the process until the valid input is generated.

\subsection{Prompt Generation for Test Generator with Mutation Rule}
\label{subsec_approach_Test_Prompt_Generation}

Based on the valid input in the previous section, {\tool} then leverages LLM to produce the test generator together with the mutation rule.
As demonstrated in Figure \ref{fig:in-context-learning} \ding{176}, the test generator is a code snippet that can generate a batch of unusual inputs, while the mutation rule is the natural language described operation for mutating the valid inputs which automatically output by LLM based on our prompt and serves as the reasoning chain for producing the test generator. Note that the mutation rule here is output by LLM. 

Each time when a test generator is produced, we can obtain a batch of automatically generated unusual text inputs, and will input them into the text widgets to check whether they have successfully made the mobile app crash. 
This test execution feedback (in Section \ref{subsubsec_approach_Feedback_Extraction}) will be incorporated in the prompt for querying the LLM which can enable it more familiar with how the mutation works and potentially produce more diversified outcomes.
We also include the inferred constraints in the previous section in the prompt (in Section \ref{subsubsec_approach_Constraint_Extraction}), since the natural language described explanation would facilitate the LLM in producing effective mutation rules, for example, the inferred constraint is that the input should be in pure text (without special characters) and the LLM would try to insert certain characters to violate the constraint.


\subsubsection{\textbf{Inferred Constraints and Valid Input Extraction}}
\label{subsubsec_approach_Constraint_Extraction}

We have obtained the inferred constraints and valid input from the output of the LLM in the previous section, here we extract this information from the output message and will input it into the LLM in this section. 
We design a flexible keyword matching method to automatically extract the description between the terms like `constraints' and `the input' and treat it as the inferred constraints, and extract the description after the terms like `input is' and treat it as the valid input, as demonstrated in Figure \ref{fig:in-context-learning} \ding{173}.

\subsubsection{\textbf{Test Execution Feedback Extraction}}
\label{subsubsec_approach_Feedback_Extraction}

After generating the unusual text inputs, we input them into the mobile app and check whether they can successfully trigger the app crash. 
This test execution information will be inputted into the LLM to generate more effective and diversified text inputs. 
We use the real buggy text inputs and the other unusual inputs (which don't trigger bugs) to prompt LLM in the follow-up generation.
The former can remind the LLM to avoid generating duplicate ones, while the latter aims at telling the LLM to consider other mutation rules. 
 
Besides, we also associate the mutation rules with the text input to enable the LLM to better capture its semantic meaning.
As shown in Figure \ref{fig:in-context-learning} \ding{176}, we extract the content between the keywords ``Mutation rule'' and ``Test generator'' as mutation rules.



\subsubsection{\textbf{Prompt Generation}}
\label{subsubsec_approach_Test_case_Prompt_Generation}

With the extracted information, we design linguistic patterns of the prompt for generating the test generator and mutation rules. 
As shown in Figure \ref{fig:in-context-learning} \ding{175}, the prompt includes four kinds of information, namely inferred constraints, valid input, text execution feedback, and question.
The first three kinds of information are mainly based on the extracted information as described above, and we also add some background illustrations to let the LLM better understand the task, like the inferred constraint in Figure \ref{fig:in-context-learning} \ding{175}.
For the question, we first ask the LLM to generate the mutation rule for the valid input, then let it produce a test generator following the mutation rule. 
Due to the robustness of LLM, the generated prompt sentence does not need to follow the grammar completely.

\subsection{Enriching Prompt with Examples}
\label{subsec_approach_Prompt-tuning}
It is usually difficult for LLM to perform well on domain-specific tasks as ours, and a common practice would be employing the in-context learning schema to boost the performance. 
It provides the LLM with examples to demonstrate what the instruction is, which enables the LLM better understand the task. 
Following the schema, along with the prompt for the test generator as described in Section \ref{subsec_approach_Test_Prompt_Generation}, we additionally provide the LLM with examples of the unusual inputs. 
To achieve this, we first build a basic example dataset of buggy inputs (which truly trigger the crash) from the issue reports of open-source mobile apps, and continuously enlarge it with the running records during the testing process (in Section \ref{subsubsec_approach_dataset_construction}). 
Based on the example dataset, we design a retrieval-based example selection method (in Section \ref{subsubsec_approach_Retrieval}) to choose the most suitable examples in terms of an input widget, which further enables the LLM to learn with pertinence. 


\subsubsection{\textbf{Example Dataset Construction}}
\label{subsubsec_approach_dataset_construction}

We collect the buggy text inputs from GitHub and continuously build an example dataset that serves as the basis for in-context learning. 
For each data instance, as demonstrated in \ref{fig:in-context-learning} \ding{174}, it records the buggy text inputs and the mutation rules which facilitate the LLM understanding of how the buggy inputs come from. 
It also includes the context information of the input widgets which provides the background information of the buggy inputs, and enables us to select the most suitable examples when querying the LLM.

\textbf{Mining buggy text inputs from GitHub.} 
First, we automatically crawl the issue reports and pull requests from the Android mobile apps in GitHub (updated before September 2022). 
Then we use keyword matching to filter these related to the text inputs (e.g., EditText) and have triggered crashes. 
Following that, we then employ manual checking to further determine whether there is a crash triggered by the buggy text inputs by running the app.
In this way, we obtain 50 unusual inputs and store them in the example dataset (There is no overlap with the evaluation datasets.).
We then extract the context information of the input widget with the method in Section \ref{subsubsec_approach_Context_Extraction}, and store it together with the unusual input.
Note that, since these buggy inputs don't associate with the mutation rules, we set them as null.


\textbf{Enlarging the dataset with buggy text inputs during testing.}
We enrich the example dataset with the newly emerged unusual text inputs which truly trigger bugs during {\tool} runs on various apps. 
Specifically, for each generated unusual text input, after running it in the mobile apps, we put the ones which trigger crashes into the example dataset. 
We also add their associated mutation rules generated by the LLM, as well as the context information extracted in Section \ref{subsubsec_approach_Context_Extraction}. 


\subsubsection{\textbf{Retrieval-based Example Selection and In-context Learning}}
\label{subsubsec_approach_Retrieval}

Examples can provide intuitive guidance to the LLM in accomplishing a task, yet excessive examples might mislead the LLM and cause the performance to decline. 
Therefore, we design a retrieval-based example selection method to choose the most suitable examples (i.e., most similar to the input widgets) for LLM.

In detail, the similarity comparison is based on the context information of the input widgets. 
We use Word2Vec (Lightweight word embedding method) \cite{Word2Vec} to encode the context information of each input widget into a 300-dimensional sentence embedding, and calculate the cosine similarity between the input widget and each data instance in the example dataset. 
We choose the top-K data instance with the highest similarity score, and set K as 5 empirically. 

The selected data instances (i.e., examples) will be provided to the LLM in the format of context information, mutation rule, and buggy text input, as demonstrated in Figure \ref{fig:in-context-learning} \ding{174}.

\subsection{Implementation}
\label{subsec_approach_Implement}
We implement {\tool} based on the ChatGPT which is released on the OpenAI website\footnote{\url{https://beta.openai.com/docs/models/chatgpt}}. 
It obtains the view hierarchy file of the current GUI page through UIAutomator~\cite{uiautomator} to extract context information of the input widgets.
{\tool} can be integrated by replacing the text input generation module of the automated GUI testing tool, which automatically extracts the context information and generates the unusual inputs.

%% file: figure/in-context-learning.tex
\begin{figure}[t!]
\centering
\vspace{0.1in}
\includegraphics[width=8.6cm]{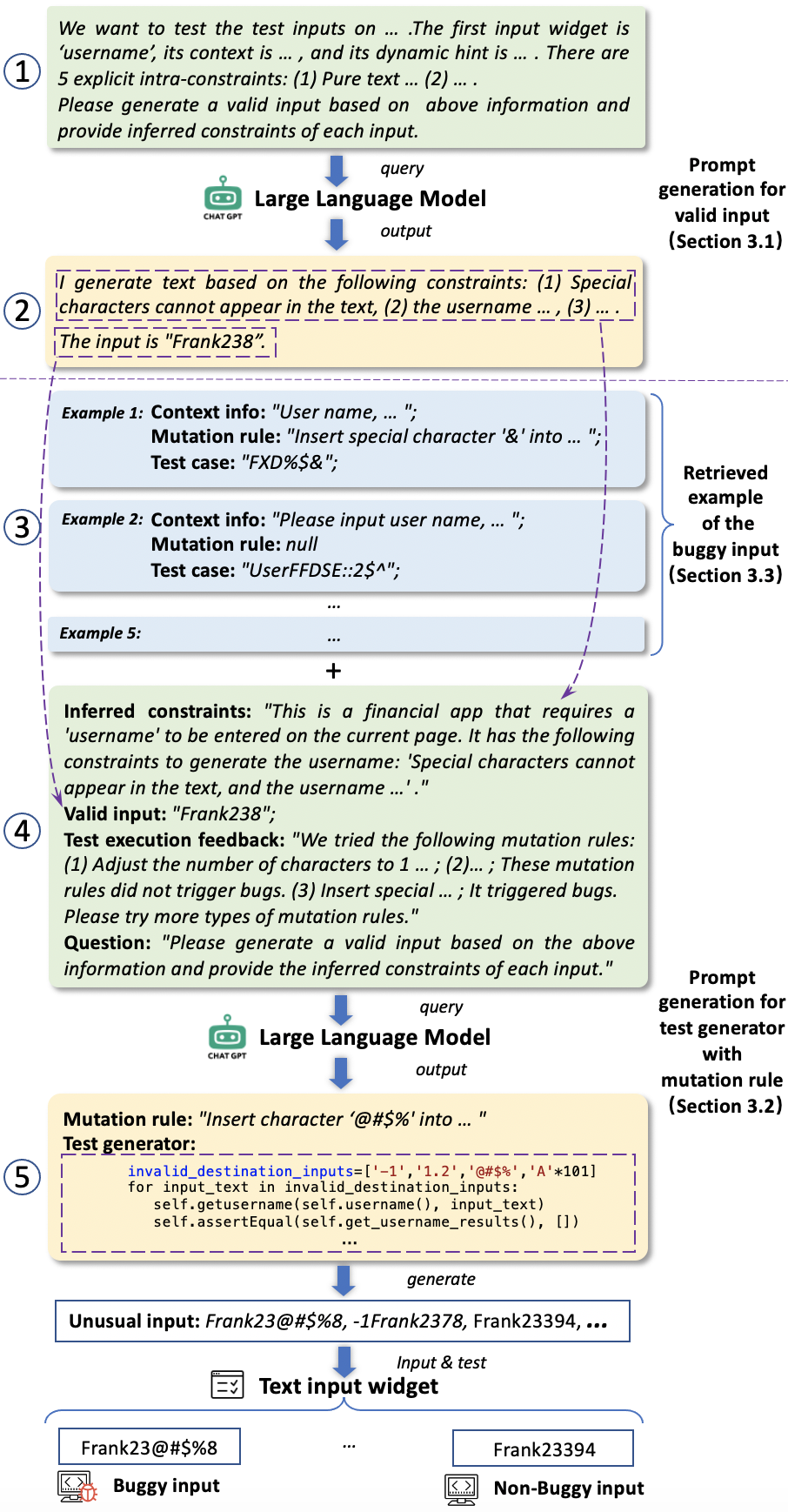}
\caption{Example of how {\tool} works. }
\label{fig:in-context-learning}
\vspace{-0.15in}
\end{figure}

%% file: tab/approach-rule.tex
\begin{table*}
\renewcommand\arraystretch{1} 
\caption{The example of extracted information and linguistic patterns of prompts for Module 1. }
\vspace{-0.1in}
\label{tab:approach-rule}
\centering
\scriptsize
\begin{center}
\begin{tabular}{m{0.3cm}<{\centering} | m{2.4cm}<{} | m{7.7cm}<{} | m{6.0cm}<{}}
\toprule
\multicolumn{4}{c}{\textbf{Extracted information}} \\ 
\toprule
\textbf{Id} & \textbf{Attribute} & \textbf{Description}  & \textbf{Examples} \\
\midrule
I1 & AppName & The name of testing app & AppName = ``Wallet'' \\
I2 & PageName & Activity name of the current GUI page & PageName = ``User'' \\
I3 & InputWidget & The text input widget(s) denoted with the textual related fields & InputWidget = ``Please input user name'' \\
I4 & NearbyWidget & Nearby widgets denoted with their textual related fields & NearbyWidget = ``your income: [SEP] \$ '' \\
\midrule
I5 & DynamicHint & Feedbacks in terms of an incorrect input & DynamicHint = ``password should contain letters'' \\
\midrule
I6 & CandidateConstraints & Candidate constraints within or among widget(s) summarized in pilot study, organized into intra-constraint(explicit), intra-constraint(implicit), and inter-constraint
& CandidateConstraints = ``intra-constraints(explicit): (1) Pure text (without special characters)  ... ''  \\
\toprule
\multicolumn{4}{c}{\textbf{Linguistic patterns of prompts}} \\ 
\toprule
\textbf{Id} & \textbf{Target} & \textbf{Pattern}  & \textbf{Examples} \\
\midrule
P1 & Provide context information of the text input widgets & We want to test the text input widgets on $\langle PageName \rangle $ page of $\langle AppName \rangle $ app which has $\langle \#NumOfInputWidget \rangle $ text inputs. The first input widget is $\langle InputWidget \rangle $ , its context is $\langle InputWidget \rangle $ , and its dynamic hint is $\langle DynamicHint \rangle $ . The second input ... . & We want to test the text input widgets on User page of Wallet app which has 3 text inputs. The first input widget is `username', its context is `Welcome to ...', and its dynamic hint is `Username already in use'. ... \\
\midrule
P2 & Provide candidate constraints & There are 5 explicit intra-constraints: $\langle intra-constraint (explicit) \rangle $ ; 5 implicit intra-constraints: $\langle intra-constraint (implicit) \rangle $ ; 7 inter-constraints: $\langle inter-constraint \rangle $ & There are 5 explicit intra-constraints: (1) Pure text ... ; 5 implicit intra-constraints: (1) Limited string length ...; 7 inter-constraints: (1) Departure and Arrival ... 
\\
\midrule
P3 & Query LLM & \multicolumn{2}{c}{Please generate a valid input based on the above information and provide the inferred constraints of each input.}\\
\bottomrule
\end{tabular}
\end{center}
\vspace{-0.15in}
\end{table*}

%% file: sec/experiment.tex
\section{Experiment Design}
\label{sec_Effectiveness}


\subsection{Research Questions}
\label{subsec_experiment_RQ}
\begin{itemize}[leftmargin=*]
\item \textbf{RQ1: (Bugs Detection Performance)} How effective of {\tool} in detecting bugs related to text input widgets? 
\end{itemize}

For RQ1, we first present some general views of {\tool} for bug detection, and then compare it with commonly-used and state-of-the-art baseline approaches. 

\begin{itemize}[leftmargin=*]
\item \textbf{RQ2: (Ablation Study)} 
What is the contribution of the (sub-) modules of {\tool} for bug detection performance?
\end{itemize}

For RQ2, We conduct ablation experiments to evaluate the impact of each (sub-) module on the performance.

\begin{itemize}[leftmargin=*]
\item \textbf{RQ3: (Usefulness Evaluation)} How does our proposed {\tool} work in real-world situations?
\end{itemize}

For RQ3, we integrate {\tool} with the GUI testing tool to make it automatically explore the app and detect unseen input-related bugs, and issue the detected bugs to the development team.

\subsection{Experimental Setup}
\label{subsec_experiment_dataset}

\textbf{For RQ1 and RQ2,} we crawl 200 most popular open-source apps from F-Droid~\cite{F-droid}, and only keep the latest ones with at least one update after September 2022 (this ensures the utilized apps are not overlapped with the ones in Sec \ref{subsec_approach_Prompt-tuning}).
Then we collect all their issue reports on GitHub, and use keywords (e.g., EditText) to filter those related to text input. 
Finally, we obtain 126 issue reports related to 54 apps. 
Then we manually review each issue report and the mobile app, and filter it according to the following criteria: (1) the app wouldn't constantly crash on the emulator; (2) it can run all baselines; (3) UIAutomator~\cite{uiautomator} can obtain the view hierarchy file for context extraction; (4) the bug is related to text input widgets; (5) the bug can be manually reproduced for validation; (6) the app is not used in the motivational study or example dataset construction. 
Please note that we follow the name of the app to ensure that there is no overlap between the datasets.
Finally, 31 apps with 36 buggy text inputs remain for further experiments.

We measure the bug detection rate, i.e., the ratio of successfully triggered crashes in terms of all the experimental crashes (i.e., buggy inputs), which is a widely used metric for evaluating GUI testing~\cite{liu2017automatic,he2020textexerciser,arnatovich2018mobolic}.
Specifically, with the generated unusual input, we design an automated test script to input it into the text input widgets, and automatically run the ``submit'' operation to check whether a crash occurs.
If no, use the script to go back to GUI page with the input widget if necessary, and try the next generated unusual input.
As long as a crash is triggered for a text input widget, we treat it as successful bug detection and will stop the generation for this widget. 
Note that our generated unusual input is not necessarily the same as the one provided in the issue report, e.g., -18 vs. -20, as long as a crash is triggered after entering the unusual inputs, we treat it as a successful crash detection. 

For a fair comparison with other approaches, we employ two experimental settings, i.e., 30 attempts (30 unusual inputs) and 30 minutes.
We record the bug detection rate under each setting (denoted as  ``Bug (\%)'' in Table \ref{tab:RQ1-result} to Table \ref{RQ2-3-example}), and also record the actual number of attempts (denoted as  ``Attempt (\#)'') and the actual running time (denoted as ``Min (\#)'') when the crash occurs to fully understanding the performance.

 




\textbf{For RQ3,} we further evaluate the usefulness of {\tool} in detecting unseen crash bugs related to text input. 
A total of 131 apps have been retained. 
We run Ape~\cite{gu2019practical} (a commonly-used automated GUI testing tool) integrated with {\tool}, for exploring the mobile apps and getting the view hierarchy file of each GUI page. 
We use the same configurations as the previous experiments. 
Once a crash related to text input is spotted, we create an issue report by describing the bug, and report them to the app development team through the issue reporting system or email.

\subsection{Baselines}
\label{subsec_experiment_baseline}

Since there are hardly any existing approaches for the unusual input generation of mobile apps, we employ 18 baselines from various aspects to provide a thorough comparison.

First, we directly utilize \textit{ChatGPT}~\cite{schulman2022chatgpt} as the baseline.
We provide the context information of the text input widgets (as described in Table \ref{tab:approach-rule} P1), and ask it to generate inputs that can make app crash.

Fuzzing testing and mutation testing can be promising techniques for generating invalid inputs, and we apply several related baselines.
Feldt et al. \cite{feldt2013finding} proposed a testing framework called \textit{GoldTest}, which generates diverse test inputs for mobile apps by designing regular expressions and generation strategies. 
In 2017, they further proposed an invalid input generation method~\cite{poulding2017generating} based on probability distribution (PD) parameters and regular expressions, and we name this baseline as \textit{PDinvalid}.
Furthermore, we reuse the idea of traditional random-based fuzzing~\cite{chen2018systematic,liang2018fuzzing}, and develop a \textit{RandomFuzz} for generating inputs for text widgets.
In addition, based on the 50 buggy text inputs from the GitHub dataset in Section \ref{subsubsec_approach_dataset_construction}, we manually design 50 corresponding mutation rules to generate the invalid input, and name this baseline as \textit{ruleMutator}.



Furthermore, we include the string analysis methods as the baselines, i.e., \textit{OSTRICH}~\cite{chen2020decision} and \textit{Sloth} ~\cite{chen2022solving}.
They aim at generating the strings that violate the constraints (e.g., string length, concatenation, etc), which is similar to our task.
\textit{OSTRICH}'s key idea~\cite{chen2020decision} is to generate the test strings based on heuristic rules.
\textit{Sloth}~\cite{chen2022solving} proposes to exploit succinct alternating finite-state automata as concise symbolic representations of string constraints.


There are constraint-based methods, i.e.,  \textit{Mobolic}~\cite{arnatovich2018mobolic} and \textit{TextExerciser}~\cite{he2020textexerciser}, which can generate diversified inputs for testing the app. 
For example, \textit{TextExerciser} utilizes the dynamic hints to guide it in producing the inputs.

We also employ two methods (\textit{RNNInput} ~\cite{liu2017automatic} and \textit{QTypist} ~\cite{liu2022fill}) which aim at generating valid inputs for passing the GUI page.
In addition, we use the automated GUI testing tools, i.e., \textit{Stoat}~\cite{su2017guided}, \textit{Droidbot}~\cite{li2017droidbot}, \textit{Ape}~\cite{gu2019practical}, \textit{Fastbot}~\cite{cai2020fastbot}, \textit{ComboDroid}~\cite{wang2020combodroid}, \textit{TimeMachine}~\cite{dong2020time}, \textit{Humanoid}~\cite{li2019humanoid}, \textit{Q-testing}~\cite{pan2020reinforcement}, which can produce inputs randomly or following rules to make app running automatically.

We design the script for each baseline to ensure that it can reach the GUI page with the text input widget, and run them in the same experimental environment (Android x64) to mitigate potential bias.

%% file: sec/result.tex
\section{Results and Analysis}
\label{sec_results}

\subsection{Bugs Detection Performance (RQ1)}
\label{sec_results_RQ1}

Table \ref{tab:RQ1-result} presents the bug detection performance of {\tool}. 
With the unusual inputs generated by {\tool}, the bug detection rate is 0.78 (within 30 minutes), indicating 78\% (28/36) of the bugs can be detected. 
In addition, the bugs can be detected with an average of 13.52 attempts, and the average bug detection time is 9.64 minutes, which is acceptable. 
This indicates the effectiveness of our approach in generating unusual inputs for testing the app, and facilitating the uncovering of bugs related to input widgets.

Figure \ref{fig:good-case} demonstrates examples of {\tool}'s generated unusual inputs and the inputs that truly trigger the crash.
We can see that our proposed approach can generate quite diversified inputs which mutate the valid input from different aspects, e.g., for the price in the first example which should be a non-negative value, the generated unusual inputs range from negative values and decimals to various kinds of character strings. 
Furthermore, it is good at capturing the contextual semantic information of the input widgets and their associated constraints, and generating the violations accordingly. 
For example, for the minimum and maximum price in the first example, it generates the unusual inputs with the minimum larger than the maximum, and successfully triggers the crash. 



We further analyze the bugs that could not be detected by our approach. 
A common feature is that they need to be triggered under specific settings, e.g., only under the user-defined setting, the input can trigger the crash, in the environment we tested, it may not have been possible to trigger a crash due to the lack of user-defined settings in advance. 
We have manually compared the unusual inputs generated by our approach with the ones in the issue reports. 
We find in all cases, {\tool} can generate the satisfied buggy inputs within 30 attempts and 30 minutes, which further indicates its effectiveness. 


\textbf{Performance comparison with baselines.} 
\label{sec_results_RQ1_1}
Table \ref{tab:RQ1-result} also shows the performance comparison with the baselines. 
We can see that our proposed {\tool} is much better than the baselines, i.e., 136\% (0.78 vs. 0.33) higher in bug detection rate (within 30 minutes) compared with the best baseline TextExerciser.
This further indicates the advantages of our approach.
Nevertheless, the TextExerciser can only utilize the dynamic hints in input generation which covers a small portion of all situations, i.e., a large number of input widgets donot involve such feedback.


\input{tab/RQ1-result}

\input{figure/good-case}

Without our elaborate design, the raw ChatGPT demonstrates poor performance, which further indicates the necessity of our approach. 
In addition, the string analysis methods, which are designed specifically for string constraints, would fail to work for mobile apps. 
In addition, since the input widgets of mobile apps are more diversified (as shown in Section \ref{sec_motivation_types_UIs}) compared with the string, the heuristic analysis or finite-state automata techniques in the string analysis methods might be ineffective for our task.
The baselines for automated GUI testing or valid text input generation are even worse, since their main focus is to increase the coverage through generating valid inputs.
This further implies the value of our approach for targeting this unexplored task. 






\subsection{Ablation Study (RQ2)}
\label{sec_results_RQ2}

\subsubsection{\textbf{Contribution of Modules}}
\label{subsub_results_RQ2_Main}
Table \ref{tab:RQ2-1-compents} shows the performance of {\tool} and its 2 variants respectively removing the first and third module. 
In detail, 
for \textit{{\tool} w/o validInput} (i.e., without Module 1), we provide the information related to the input widgets (as Table \ref{tab:approach-rule} P1) to the LLM in Module 2 and set other information from Module 1 as ``null''.
For \textit{{\tool} w/o enrichExamples} (i.e., without Module 3), we set the examples from Module 3 as ``null'' when querying the LLM. 
Note that, since Module 2 is for generating the unusual inputs which is indispensable for this task, hence we donot experiment with this variant.

We can see that {\tool}'s bug detection performance is much higher than all other variants, indicating the necessity of the designed modules and the advantage of our approach.

Compared with {\tool}, \textit{{\tool} w/o validInput} results in the largest performance decline, i.e., 50\% drop (0.39 vs. 0.78) in bug detection rate within 30 minutes.
This further indicates that the generated valid inputs and inferred constraints in Module 1 can help LLM understand what the correct input looks like and generate the violated ones.

\textit{{\tool} w/o enrichExamples} also undergoes a big performance decrease, i.e., 32\% (0.53 vs. 0.78) in bug detection rate within 30 minutes, and the average testing time increases by 109\% (9.64 vs. 20.15).
This might be because without the examples, the LLM would spend more time understanding user intention and criteria for what kinds of answers are wanted. 

\input{tab/RQ2-1-compents}


\subsubsection{\textbf{Contribution of Sub-modules}}
\label{subsub_results_RQ2_Main_Example}
Table \ref{tab:RQ2-2-subcompents} further demonstrates the performance of {\tool} and its 5 variants. We remove each sub-module of the {\tool} in Figure \ref{fig:overview} separately, i.e., inferred constraint, mutation rule, text execution feedback, test generator and retrieved examples of buggy input. 
For removing the test generator, we directly let the LLM generate the unusual inputs, and for removing retrieved examples, we use the random selection method. 
For other variants, we set the removed content as ``null''.


\input{tab/RQ2-2-subcompents}

\input{tab/RQ2-3-example}

The experimental results demonstrate that removing any of the sub-modules would result in a noticeable performance decline, indicating the necessity and effectiveness of the designed sub-modules. 

Removing the mutation rules (\textit{{\tool} w/o-mutateRule}) have the greatest impact on the performance, reducing the bug detection rate by 50\% (0.36 vs. 0.72 within 30 attempts).
Remember that, {\tool} first lets the LLM to generate the mutation rules (how to mutate the valid inputs), then asks it to produce the test generator following the mutation rule.
With the generated mutation rules serving as the reasoning chain, the unusual input generation can be more effective, which further proves the usefulness of our design. 

We also notice that, when removing the test generator (\textit{{\tool} w/o-generator}), the bug detection rate does not drop much (0.72 vs. 0.61) when considering 30 attempts, yet it declines a lot (0.78 vs. 0.36) when considering 30 minutes of testing time. 
This is because our proposed approach lets the LLM produce the test generator which can yield a batch of unusual inputs. This means interacting with the LLM once can generate multiple outcomes. However, if asking the LLM to directly generates unusual inputs (i.e., \textit{{\tool} w/o-generator}), it requires interacting with LLM frequently, and could be quite inefficient.
This further demonstrates we formulate the problem as producing the test generator task is efficient and valuable. 

In addition, randomly selecting the examples (\textit{{\tool} w/o-retriExample}) would also largely influence the performance, and decrease the bug detection rate by 22\% (0.56 vs. 0.72 within 30 attempts).
This indicates that by providing similar examples, the LLM can quickly think out what should the unusual inputs look like. 
Nevertheless, we can see that, compared with the variant without enriched examples in prompt (Table \ref{tab:RQ2-1-compents}), the randomly selected examples do take effect (0.47 vs 0.56 in bug detection rate within 30 attempts), which further indicates the demonstration can facilitate the LLM in producing the required output.



\subsubsection{\textbf{Influence of Different Number of Examples}}
\label{subsub_results_RQ2_Main_Example_data}

Table \ref{RQ2-3-example} demonstrates the performance under the different number of examples provided in the prompt.


We can see that the number of detected bugs increases with more examples, reaching the highest bug detection rate with 5 examples.  
And after that, the performance would gradually decrease even increasing the examples. 
This indicates that too few or too many examples would both damage the performance, because of the tiny information or the noise in the provided examples. 



\input{tab/usefulness}

\subsection{Usefulness Evaluation (RQ3)}
\label{sec_results_RQ3}
Table \ref{tab:RQ3-Usefulness} shows all bugs spotted by Ape integrated with our {\tool}, and more detailed information on detected bugs can be seen in our website.
For the 131 apps, {\tool} detects 43 bugs in 32 apps, of which 37 are newly-detected bugs. 
Furthermore, these new bugs are not detected by the Ape without {\tool}. 

We submit these 37 bugs to the development team, and 28 of them have been fixed/confirmed so far (21 fixed and 7 confirmed), while the remaining are still pending (none of them is rejected). This further indicates the effectiveness and usefulness of our proposed {\tool} in bug detection.

When confirming and fixing the bugs, some Android app developers express thanks such as ``\textit{Very nice! You find an invalid input we thought was too insignificant to cause crashes.}''(i.e., Ipsos).
Furthermore, some developers also express their thought about the buggy text input ``\textit{Handling different inputs can be tricky, and I admit we couldn't test for every possible scenario. It has given me a fresh appreciation for the complexity of user inputs and the potential bugs they can introduce. }''(i.e., DRBUs). Some developers also present valuable suggestions to facilitate the further improvement of {\tool}.
For example, some of them hope that we can find the patterns of these bugs and design repair methods. 



%% file: tab/RQ1-result.tex
\begin{table}[!t]
\vspace{0.05in}
\renewcommand\arraystretch{1} 
\caption{Result of bugs detection performance. (RQ1)}
\vspace{-0.05in}
\label{tab:RQ1-result}
\centering
\footnotesize
\begin{tabular}{p{1.8cm}<{\centering} | p{1.1cm}<{\centering} | p{1.2cm}<{\centering} || p{1.1cm}<{\centering} | p{1.2cm}<{\centering}}
\toprule
\multirow{2}*{\textbf{Method}} & \multicolumn{2}{c||}{\textbf{ Setting 1 (30 attempts)}} & \multicolumn{2}{c}{\textbf{Setting 2 (30 minutes)}}  \cr 
 &  \textbf{Bug(\%)} & \textbf{Attempt(\#)} & \textbf{Bug(\%)} & \textbf{Min(\#)} \\
\midrule 
\textbf{InputBlaster} & \textbf{0.72}  & \textbf{13.52}  & \textbf{0.78}  & \textbf{9.64} \\
\midrule
ChatGPT & 0.25  & 25.91  & 0.28  & 23.28 \\
\midrule
\multicolumn{5}{c}{\textbf{Mutation or fuzzing methods}}\\
\midrule
GoldTest & 0.08  & 29.22  & 0.08  & 28.73 \\
PDinvalid & 0.19  & 28.65  & 0.19  & 22.73 \\
RandomFuzz & 0.25  & 22.31  & 0.25  & 21.55 \\
ruleMutator & 0.28  & 21.42  & 0.28  & 20.53 \\
\midrule
\multicolumn{5}{c}{\textbf{String analysis methods}}\\
\midrule
Sloth & 0.25  & 23.61  & 0.25  & 22.61 \\
OSTRICH & 0.22  & 24.14  & 0.22  & 23.41 \\
\midrule
\multicolumn{5}{c}{\textbf{Constraint-based methods}}\\
\midrule
Mobolic & 0.17  & 25.83  & 0.17  & 25.09 \\
TextExerciser & 0.31  & 22.11  & 0.33  & 20.18 \\
\midrule
\multicolumn{5}{c}{\textbf{Valid input generation methods}}\\
\midrule
RNNInput & 0.06  & 28.67  & 0.06  & 28.64 \\
QTypist & 0.08  & 27.78  & 0.11  & 27.31 \\
\midrule
\multicolumn{5}{c}{\textbf{Automated GUI testing methods}}\\
\midrule
Ape & 0.08  & 28.11  & 0.11  & 26.88 \\
DroidBot & 0.06  & 28.39  & 0.06  & 28.34 \\
Stoat & 0.08  & 27.94  & 0.08  & 27.58 \\
TimeMachine & 0.11  & 26.92  & 0.11  & 26.69 \\
ComboDroid & 0.14  & 26.11  & 0.14  & 25.85 \\
Q-testing & 0.11  & 27.06  & 0.11  & 26.70 \\
Humanoid & 0.11  & 26.92  & 0.14  & 25.85 \\
\bottomrule
\end{tabular}
\vspace{0.05in}
\begin{tablenotes}
\scriptsize
\item \textbf{\textit{Notes:}} ``Bug (\%)'' is the average bug detecting rate, ``Attempt (\#)'' is the average number of unusual inputs before triggering the crash, ``Min (\#)'' is the average running time (minutes) before triggering the crash. 
\end{tablenotes}
\vspace{-0.15in}
\end{table}

%% file: figure/good-case.tex
\begin{figure}[htb]
\centering
\vspace{0.05in}
\includegraphics[width=8.3cm]{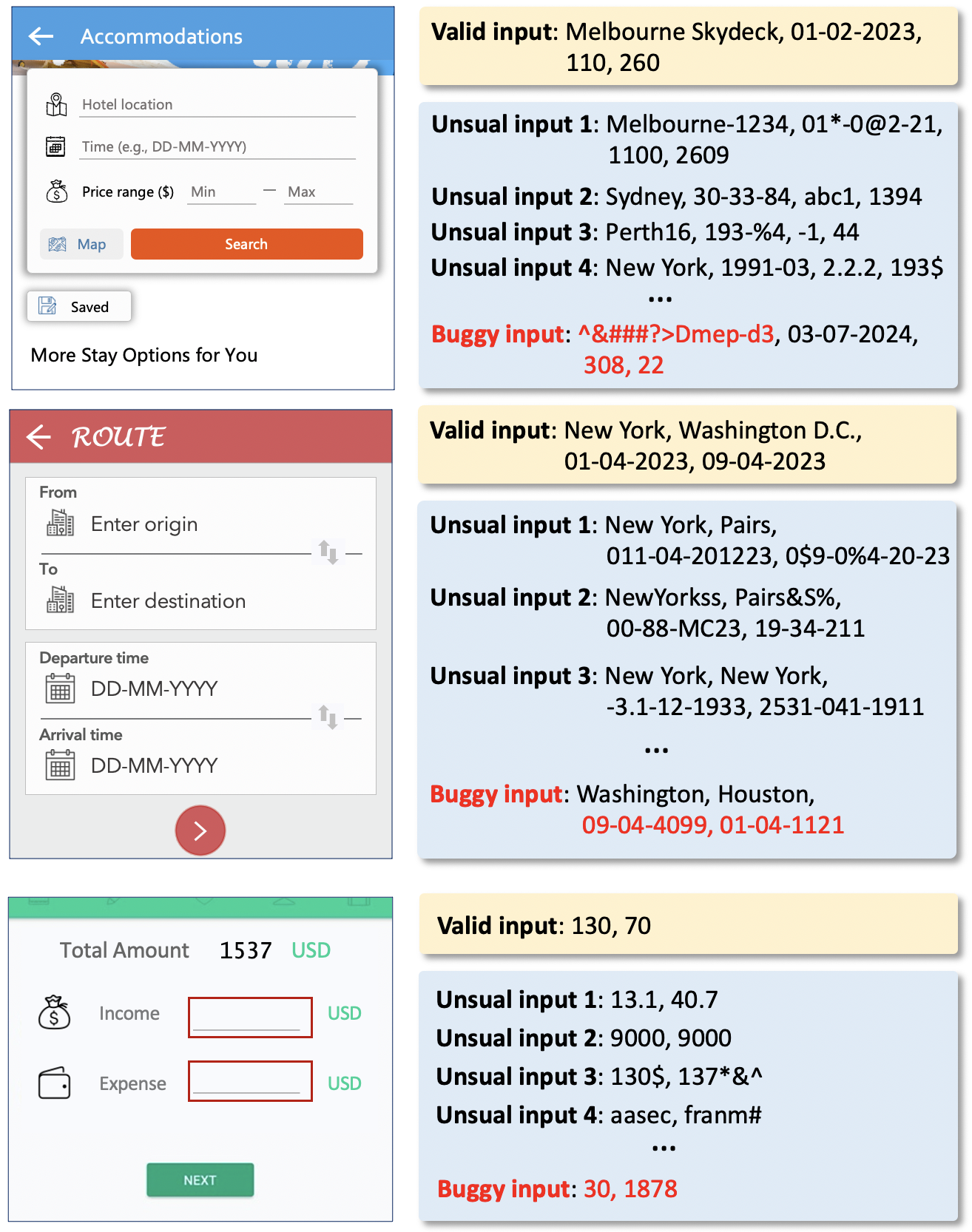}
\caption{Example of {\tool}'s output.}
\label{fig:good-case}
\vspace{-0.15in}
\end{figure}

%% file: tab/RQ2-1-compents.tex
\begin{table}[!t]
\renewcommand\arraystretch{1} 
\caption{Contribution of different modules (RQ2)}
\vspace{-0.05in}
\label{tab:RQ2-1-compents}
\centering
\footnotesize
\begin{tabular}{p{2.6cm}<{\centering} | p{0.9cm}<{\centering} | p{1.3cm}<{\centering} || p{0.9cm}<{\centering} | p{0.9cm}<{\centering}}
\toprule
\multirow{2}*{\textbf{Method}} & \multicolumn{2}{c||}{\textbf{30 attempts}} & \multicolumn{2}{c}{\textbf{30 minutes}}  \cr 
 &  \textbf{Bug(\%)} & \textbf{Attempt(\#)} & \textbf{Bug(\%)} & \textbf{Min(\#)} \\
 \midrule
\textbf{InputBlaster (Base)} & \textbf{0.72}  & \textbf{13.52}  & \textbf{0.78}  & \textbf{9.64} \\
\midrule
\textit{w/o Module 1 }& 0.31 & 22.75 & 0.39 & 19.15 \\
\textit{w/o Module 3 }& 0.47 & 22.19 & 0.53 & 20.15 \\
\bottomrule
\end{tabular}
\vspace{0.05in}
\begin{tablenotes}
\scriptsize
\item \textbf{\textit{Notes:}} The two variants respectively denote {\tool} removing module 1 (valid input generation) and module 3 (enriched examples in prompt).
\end{tablenotes}
\vspace{-0.1in}
\end{table}

%% file: tab/RQ2-2-subcompents.tex
\begin{table}[!t]
\renewcommand\arraystretch{1} 
\caption{Contribution of different sub-modules (RQ2)}
\vspace{-0.05in}
\label{tab:RQ2-2-subcompents}
\centering
\footnotesize
\begin{tabular}{p{2.6cm}<{\centering} | p{0.9cm}<{\centering} | p{1.3cm}<{\centering} || p{0.9cm}<{\centering} | p{0.9cm}<{\centering}}
\toprule
\multirow{2}*{\textbf{Method}} & \multicolumn{2}{c||}{\textbf{ 30 attempts}} & \multicolumn{2}{c}{\textbf{30 minutes}}  \cr 
 &  \textbf{Bug(\%)} & \textbf{Attempt(\#)} & \textbf{Bug(\%)} & \textbf{Min(\#)} \\
 \midrule
\textbf{InputBlaster (Base)} & \textbf{0.72}  & \textbf{13.52}  & \textbf{0.78}  & \textbf{9.64} \\
\midrule
\textit{w/o inferCons} & 0.53 & 19.94 & 0.56 & 15.11 \\
\textit{w/o mutateRule} & 0.36 & 21.31 & 0.42 & 20.71 \\
\textit{w/o feedback} & 0.58 & 16.64 & 0.58 & 14.40 \\
\textit{w/o generator} & 0.61 & 16.86 & 0.36 & 24.37 \\
\textit{w/o retriExample} & 0.56 & 19.11 & 0.56 & 23.44 \\
\bottomrule
\end{tabular}
\vspace{0.05in}
\begin{tablenotes}
\scriptsize
\item \textbf{\textit{Notes:}} The five variants respectively denote {\tool} removing inferred constraint, mutation rule, test execution feedback, test generator, retrieved examples of buggy input.  
\end{tablenotes}
\vspace{-0.1in}
\end{table}

%% file: tab/RQ2-3-example.tex
\begin{table}[!t]
\renewcommand\arraystretch{1} 
\caption{Result of different number of examples. (RQ2)}
\vspace{-0.05in}
\label{RQ2-3-example}
\centering
\footnotesize
\begin{tabular}{p{1.7cm}<{\centering} | p{1.1cm}<{\centering} | p{1.2cm}<{\centering} || p{1.1cm}<{\centering} | p{1.2cm}<{\centering}}
\toprule
\multirow{2}*{\textbf{Exmample (\#)}} & \multicolumn{2}{c||}{\textbf{ Setting 1 (30 attempts)}} & \multicolumn{2}{c}{\textbf{Setting 2 (30 minutes)}}  \cr 
 &  \textbf{Bug(\%)} & \textbf{Attempt(\#)} & \textbf{Bug(\%)} & \textbf{Min(\#)} \\
\midrule
1 & 0.50 & 20.19 & 0.50 & 22.98 \\
2 & 0.53 & 19.36 & 0.56 & 18.31 \\
3 & 0.61 & 16.86 & 0.64 & 14.93 \\
4 & 0.69 & 14.36 & 0.69 & 11.14 \\
\textbf{5(InputBlaster)} & \textbf{0.72} & \textbf{13.52} & \textbf{0.78} & \textbf{9.64} \\
6 & 0.61 & 16.86 & 0.58 & 15.48 \\
7 & 0.53 & 19.69 & 0.53 & 17.15 \\
8 & 0.44 & 21.86 & 0.42 & 20.47 \\
9 & 0.38 & 23.53 & 0.36 & 22.34 \\
10 & 0.36 & 24.36 & 0.31 & 23.81 \\
\bottomrule
\end{tabular}
\vspace{-0.15in}
\end{table}

%% file: tab/usefulness.tex
\begin{table}[!t]
\caption{Confirmed or fixed bugs. (RQ3)}
\vspace{-0.05in}
\label{tab:RQ3-Usefulness}
\centering
\footnotesize
\begin{tabular}{p{0.55cm}<{\centering} | p{1.85cm}<{\centering} | p{1.2cm}<{\centering} | p{1.2cm}<{\centering} | p{1.4cm}<{\centering}}
\toprule
\textbf{Id} & \textbf{APP Name} & \textbf{Category} & \textbf{Download} & \textbf{Status}\\
\midrule
1 & OTOMU & Music & 100M+ & fixed\\

2 & KWork & Tool & 50M+ & confirmed\\  

3 & NoxSecu & Tool  & 50M+ & fixed\\  

4 & EarnMon & Finance & 50M+ & fixed\\  

5 & RewardM & Finance & 50M+ & confirmed\\  

6 & AttaPOl & Tool & 10M+  & confirmed\\  

7 & ISAY & Commun & 10M+  & fixed \\

8 & Ipsos & Commun & 10M+  & fixed\\  

9 & MediaFire & Product & 5M+  & confirmed\\ 

10 & DRBUs & Navig & 500K+ & fixed\\   

11 & MyTransp & Travel & 500K+ & fixed\\  

12 & MMDR & Utilities & 500K+ & fixed \\  

13 & Genting & Travel & 500K+ & fixed \\ 

14 & Fair & Health & 500K+ & confirmed \\  

15 & ClassySha & Tool & 500K+  & fixed \\  

16 & Linphone & Commun & 50K+  & confirmed \\ 

17 & IvyWall & Finance & 50K+  & fixed \\ 

18 & Monefy & Finance & 50K+  & fixed \\ 

19 & Spend & Finance & 50K+  & fixed \\ 

20 & NYBA & Tool & 50K+  & fixed \\ 

21 & OneTravel & Travel & 50K+ & fixed\\  

22 & Passpor & Travel & 50K+ & fixed \\  

23 & Thatch & Travel & 50K+ & confirmed \\ 

24 & Click & Utilities & 50K+ & fixed \\  

25 & GGBN & Utilities & 50K+  & fixed \\  

26 & Vived & Utilities & 50K+  & fixed \\ 

27 & Bizbazar & Finance & 50K+  & fixed \\ 

28 & Flowx & Tool & 50K+  & fixed \\

\bottomrule

\hline
\end{tabular}
\vspace{-0.15in}
\end{table}

%% file: sec/discussion.tex
\section{Discussion and Threats to Validity}
\label{sec_discussion}

\subsection{Generality Across Platforms}


The primary idea of {\tool} is to generate unusual inputs for text widgets with the context information when running the apps.
Although we only experiment with Android mobile apps, since other platforms have these similar types of information, {\tool} can be used to conduct the testing of input widgets for other platforms.
We conduct a small-scale experiment for another two popular platforms, and experiment on 10 iOS apps with 15 bugs and 10 Web apps with 18 bugs, with details on our website.
Results show that {\tool}'s bug detection rate is 80\% for iOS apps and 78\% for Web apps within 30 minutes testing time.
This further demonstrates the generality and usefulness of {\tool}, and we will conduct more thorough experiments in the future.

\subsection{Threats of Validity}
\label{subsec_Validity}

The first threat concerns the representativeness of the experimental apps. We have selected popular and active apps which can partially reduce this threat. 

The second threat relates to the baseline selection. Since there are hardly any existing approaches for the unusual input generation of mobile apps, we employ 18 approaches from various aspects for a thorough comparison. 
There are inputs generation techniques for Web apps \cite{trinh2014s3,anand2012automated,sunman2022automated,alshahwan2011automated}, yet because they need to analyze the web code which is different from mobile apps considering the different rendering mechanism, and cannot be directly applied in our task, hence we don't include them as the baselines.

The third threat is that we only focus on the crash bugs, since they cause more serious effects and can be automatically observed, and existing studies also only explore this type of bug \cite{DroidBot,li2019humanoid,pan2020reinforcement}.

The fourth threat might lie in the process of manual categorization in Section \ref{sec_motivation_types_UIs}. The process involves multiple practitioners and double-checking for the final decision. Also note that, the derived categorization is only for illustration, rather than serving as the ground truth for evaluation. 

The Fifth threat may exist in the uncertainty of LLM output results. LLM may not generate the corresponding output as expected, and we also design in-context learning and feedback mechanisms to ensure the output format and content of LLM.

Last but not least, {\tool} gradually builds the example dataset (Section \ref{subsubsec_approach_dataset_construction}) as the test goes on. 
This indicates the performance can be influenced by the testing order, e.g., when arranged in the first place, the crash could not be detected, yet when arranged after 10 apps are tested, the crash can be revealed, since the example dataset has accumulated more knowledge. 
In this paper, we use a random order of the experimental apps and would explore more in the future.



%% file: sec/related.tex
\section{Related Work}
\label{sec_related}

\textbf{Testing Related with Text Inputs.} 
There have been many automated GUI testing techniques for mobile apps \cite{machiry2013dynodroid,mao2016sapienz,gu2019practical,su2017guided,li2017droidbot,cai2020fastbot,wang2020combodroid,dong2020time,azim2013targeted,rastogi2013appsplayground,Monkey,arnatovich2016achieving}, yet they mainly focus on how to plan the exploration paths to fully cover the app activities and states. 
There are also studies \cite{liu2022fill,liu2017automatic,he2020textexerciser} that aim at generating valid inputs to pass the GUI pages and are used to enrich the automated testing tools for higher coverage. 
None of them can conduct the testing of text input widgets. 

For Web apps, SWAT~\cite{alshahwan2011automated} and  AWET~\cite{sunman2022automated} generated the unusual inputs based on the pre-defined template.
ACTEve~\cite{anand2012automated} and S3~\cite{trinh2014s3} first used symbolic execution to extract input constraints in the source code and then employ a solver to generate the inputs.
They need to analyze the web code and can't be directly applied to Android apps which have quite different rendering mechanisms. In addition, some constraints are dynamically generated (as shown in Section \ref{sec_motivation_types_UIs}), and couldn't be extracted from the source code. 

There are some string analysis methods for generating the strings that violate the constraints (e.g., string length)~\cite{chen2022solving,holik2017string,trinh2017model,liang2014dpll,li2013pass,chen2020decision,krings2020towards,day2019solving,kiezun2013hampi}.
Although they are effective for string constraints, yet the inputs of mobile apps are more diversified, and they cannot work well in our task.

\textbf{LLM for Software Engineering.} 
With the breakthrough of LLMs, studies have proposed to explore
how LLMs can be used to assist developers in a variety of tasks, such
as code generation \cite{poesia2022synchromesh,zeng2022extensive}, program repair \cite{jiang2023impact,hu2019re,nashid2023retrieval},
and code summarization \cite{zeng2022extensive,ahmed2022few}.
There is also a growing trend of applying LLM for software testing, e.g., fuzzing deep learning libraries~\cite{deng2023large}, unit test generation \cite{lemieux2023codamosa}, bug reproduction~\cite{kang2022large}, valid input generation \cite{liu2022fill}, etc, and achieves significant performance improvement. 
This work explores a different task, i.e., unusual text input generation for mobile apps, which provides new insights into how LLM can enhance the software testing practice. 

%% file: sec/conclusion.tex
\section{Conclusion}
\label{sec_conclusion}

Automated testing is crucial for helping improve app quality.
Despite the dozens of mobile app GUI testing techniques, how to automatically generate the diversified unusual text inputs for fully testing mobile apps remains a challenge. 
This paper proposes {\tool} which leverages the LLM to produce the unusual inputs together with the mutation rules which serve as the reasoning chains. It formulates the unusual inputs generation problem as a task of producing a set of test generators, each of which can yield a batch of unusual text inputs under the same mutation rule. 
The evaluation is conducted for both effectiveness and usefulness, with 136\% higher bug detection rate than the best baselines, and uncovering 37 new crashes. 

In the future, we plan to further analyze the root causes and repair strategy of these input-related bugs, and design automated bug repair methods.
